\documentclass[11pt,american,aip,floatfix,jcp,reprint,longbibliography,nofootinbib]{revtex4-1}

\usepackage{graphicx}
\usepackage{dcolumn}
\usepackage{bm}
\usepackage{hyperref}

\usepackage{amsmath}
\usepackage{amsfonts}

\DeclareMathOperator*{\lcm}{LCM}

\begin{document}

\preprint{}

\title{Graph similarity drives zeolite diffusionless transformations and intergrowth}

\author{Daniel Schwalbe-Koda}
\author{Zach Jensen}
\author{Elsa Olivetti}
\author{Rafael G\'omez-Bombarelli}
\email{rafagb@mit.edu}

\affiliation{%
 Department of Materials Science and Engineering, Massachusetts Institute of Technology, Cambridge, MA 02139
}%

\date{\today}

\begin{abstract}
\end{abstract}

\maketitle

\textbf{\footnote{Pre-peer review manuscript}
Predicting and directing polymorphic transformations is a critical challenge in zeolite synthesis\cite{Davis2002OrderedApplications,Maldonado2013,Li2015SynthesisStructures,Gallego2017AbReactions}. Although interzeolite transformations enable selective crystallization\cite{Honda2013RoleProcess.,Marler2012HydrousReview,Eliasova2015TheZeolites,Li2018}, their design lacks predictions to connect framework similarity and experimental observations. Here, computational and theoretical tools are combined to data-mine, analyze and explain interzeolite relations. It is observed that building units are weak predictors of topology interconversion and insufficient to explain intergrowth. By introducing a supercell-invariant metric that compares crystal structures using graph theory, we show that topotactic and reconstructive (diffusionless) transformations occur only between graph-similar pairs. Furthermore, all known instances of intergrowth occur between either structurally-similar or graph-similar frameworks. Backed with exhaustive literature results, we identify promising pairs for realizing novel diffusionless transformations and intergrowth. Hundreds of low-distance pairs are identified among known zeolites, and thousands of hypothetical frameworks are connected to known zeolites counterparts. The theory opens a venue to understand and control zeolite polymorphism.
}

Traditionally, zeolites are compared according to their framework density \cite{Goel2015SynthesisAgents} or set of constituent units \cite{Baerlocher2007AtlasTypes}. It is typically understood that the crystallization of certain species during interzeolite conversion is faster when the seed and the product zeolite share composite building units (CBUs)\cite{Li2018, Honda2013RoleProcess.}. The increased stability of denser frameworks is also usually described in terms of Ostwald's rule\cite{Goel2015SynthesisAgents}. This has led to a formulation of heuristic rules such as the common-CBU hypothesis to design organic structure-directing agents (OSDAs)-free routes through hydrothermal treatments. \cite{Honda2013RoleProcess., Zhang2018Understanding}

\begin{figure*}[htb]
\includegraphics[width=\linewidth]{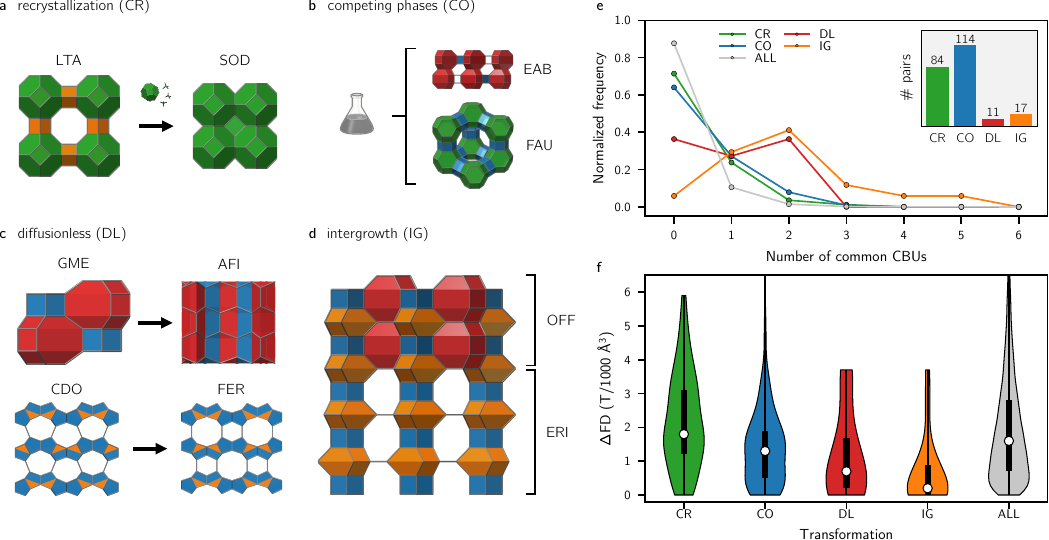}
\caption{\textbf{Types of zeolite transformations and classical explanations.} \textbf{a-d}, Diagrams of interzeolite transformations found in the literature. Colored substructures depict CBUs. \textbf{e}, Fraction of pairs experimentally observed within interzeolite transformations and their number of common CBUs. \textbf{e (inset)}, Histogram of the literature extraction. The number above the bar corresponds to the number of unique pairs found in the literature under that category. \textbf{f}, Statistical distribution of differences of framework density ($\Delta$FD) between the source and the target zeolite for each transformation. The white dot indicate the median of the distribution. The bottom and top of the thicker solid lines are the first and third quartiles, respectively, while the whiskers indicate the maximum and the minimum of the range. For \textbf{e}, \textbf{f}, the zeolite conversions are abbreviated as following: recrystallization (CR), competing phases (CO), diffusionless (DL), and intergrowth (IG). For comparison, the distributions for all pairs of known zeolites (ALL) are also shown.}
\label{fig:transformations}
\end{figure*}

Most interzeolite transformations reported in the literature can be described as recrystallization under hydrothermal treatment. Other less common types of zeolite conversion have not been as thoroughly compiled and investigated. We searched through more than 70,000 articles related to zeolites with a combination of natural-language processing and human supervision and identified 374 experimental reports of polymorphic pairwise relations. We classified these pairs into four major groups of pairwise structural relationships: recrystallization, diffusionless transformations, formation of competing phases, and intergrown phases (see Fig. \ref{fig:transformations}a-d and the Supplementary Information for full definitions).

By statistical analysis of the collected pairs, it was found that neither the framework density nor the common-CBU hypothesis by themselves explain the reported phase relations. Fig. \ref{fig:transformations}e shows the tally of transformations reported between zeolite pairs that share a given number of common CBUs. At least 35\% of recrystallization, competing and diffusionless relations have initial and final zeolites without any common CBUs. In contrast, 95\% of the unique intergrown zeolites pairs have at least one common building unit. Still, nearly 65\% of these pairs do not share the same set of CBUs (see Supp. Fig. 1). This data-driven view suggests that the common-CBU rule is not a predictor of interzeolite transitions, despite its roles on crystallization rates and topological description. Common CBUs only partially drive intergrowth.

Changes in framework densities ($\Delta$FD) do not provide significant trends for each transformation in the literature. Similarities in density are very common, as evidenced by the distribution of all pairwise density differences in Fig. \ref{fig:transformations}f. Hence, it is not a selective predictor. In recrystallization processes and competing phases, we identified a broader distribution for $\Delta$FD than in other transformations, as low-density frameworks such as FAU or LTA are usually crystallized first in hydrothermal synthesis. Their wide commercial availability also make them a popular silicon source for seed-assisted conversion of zeolites, which may skew the diversity of data available in the literature. On the other hand, more direct diffusionless transformations and intergrown zeolites tend to have smaller differences of framework density. 

\begin{figure}[tb]
\includegraphics[width=\linewidth]{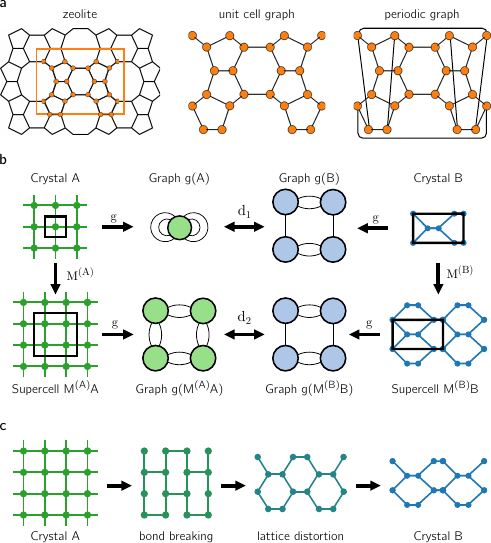}
\caption{\textbf{Graph and supercell matching.} \textbf{a}, Representation of a zeolite using a graph. The unit cell graph is modified to satisfy periodic boundary conditions by looping bonds back into the unit cell. \textbf{b}, Graph distance between different hypothetical crystal structures. The distance $d$ between crystal graphs varies with the choice of the crystallographic unit cell. In the given example, the transformation matrices are $M^{(A)} = 2 \mathbb{I}$ and $M^{(B)} = \mathbb{I}$, with $\mathbb{I}$ the identity matrix. With the choice of an appropriate metric, $d_2 \leq d_1$. \textbf{c}, Hypothetical A-B transformation. The bond breaking step removes two extra edges from the $M^{(A)} A$ crystal graph to match the $M^{(B)} B$ crystal graph, and is followed by a diffusionless transformation at constant graph.}
\label{fig:graph}
\end{figure}

The presence of one or more common CBUs or low $\Delta$FD, thus, can help rationalize results, but is not predictive of interzeolite relations. We propose that complementary, richer descriptors based on topology and structure can achieve improved explanatory and predictive power for some classes of experimental relations. Graphs are well-known representations to encode the topology of zeolites \cite{OKeeffe1996TheStructures,Foster2004,Treacy2004EnumerationGraphs,Witman2018CuttingZeolites}. Frameworks are represented as multigraphs, or crystal graphs, that label T-O-T (T = Si, Al etc.) covalent bonds as edges and capture periodic boundaries (Fig. \ref{fig:graph}a). To compare different topologies, a metric of distance between graphs is necessary. We first looked at graph isomorphism, which verifies if two graphs are equivalent up to a relabeling of their nodes\cite{Cordella2004}. Starting from the set of 245 known zeolites from the International Zeolite Association (IZA) database, we checked for the existence of crystal graph isomorphism for all 29,890 pairwise combinations of zeolite graphs. If all graph representations were distinct, no isomorphism would be detected. However, we found 14 pairs and one trio of different zeolite topologies with the same multigraph connectivity. Table \ref{tab:isomorphs} shows the isomorphic pairs identified among the known zeolites. Only four of these pairs share the exact same set of CBUs, namely CDO-FER, SFO-AFR, RSN-VSV, and AWO-UEI. All the remaining pairs have identical graph representations, but different CBUs. This suggests that graphs capture framework similarity in a distinctive way. 

\begin{table}[tb]
\caption{Pairs of known zeolites whose crystal graphs are multigraph isomorphic. Bold pairs share the same set of CBUs.}
      \begin{tabular}{cccc}
        \hline
        \textbf{CDO-FER} & \textbf{SFO-AFR} & \textbf{RSN-VSV} & \textbf{AWO-UEI}\\
        AFI-GME & AHT-ATV & CGF-HEU & JBW-NPO\\
        ABW-BCT & AWW-RTE & APC-APD & BOF-LAU\\
        MER-PHI & SBN-THO & \multicolumn{2}{c}{ACO-GIS-LTJ}\\ \hline
      \end{tabular}
      \label{tab:isomorphs}
\end{table}

Inspecting the transformations extracted from the literature, we found that three of the graph isomorphic pairs are related experimentally through diffusionless transformations: CDO-FER, GME-AFI, and APC-APD. Of those, GME-AFI and APC-APD do not have the same CBUs. Thus, Tab. \ref{tab:isomorphs} contains 27\% of the eleven known unique diffusionless zeolite conversions. This shows that graph isomorphism is much more selective than randomly guessing 11 diffusionless transformations out of 29,890 possible pairs of known zeolites. Furthermore, most of the the other graph-isomorphic zeolites are reportedly connected through other relations, experimentally-derived structural models, building schemes, and simulations (see Appendix B).

The isomorphism test provides a powerful similarity metric for zeolites, and hints at the presence of kinetic transformation channels between crystal phases\cite{Blatov2007}. The perfect equivalence between graphs ensures than both structures have the same number of atoms and bonds inside the unit cell, so there is no net bond breaking or formation. A bijection between nodes is also guaranteed: for each atom in the starting crystal, there is an equivalent atom in the final crystal with the same neighborhood. Hence, the initial and final structures are related by a local rearrangement of the atoms. The transition between the frameworks is either purely displacive (no covalent bonds are broken) or concerted (military), resembling martensitic transitions in metallurgy\cite{porter2009phase}. In contrast, hydrothermal treatments lead to the formation of an amorphous phase, essentially breaking down the topology and rebooting the crystallization process. This implies a less selective transformation if not assisted by OSDAs. Other experimental parameters may have a larger influence on the outcomes than the precrystallized reactants themselves, as in the case of competing phases.

The relationship between graph isomorphism and interzeolite conversions is illustrated with the isochemical phase transition between GME and AFI. Alberti \textit{et al.} studied this reconstructive phase transformation upon heating and identified the existence of an intermediate, ``transient'' phase with three-connected T atoms \cite{Alberti2017}. Dusselier \textit{et al}. later explained this effect using powder X-ray diffraction (XRD) pattern measurements\cite{Dusselier2017}. They proposed a mechanism of 18 T-O bonds breaking under compression of the \textit{gme} cage \cite{Dusselier2017}. Using the graph isomorphism criterion, this transition is explained through node and edge equivalences. Although the kinetic process involves bond breaking, the net number of bonds formed per unit cell is zero. Such a mechanism between the structures can be visualized by interpolating the equivalent atomic positions of each crystal. Supp. Fig. 2 depicts the evolution of the GME-AFI transformation, compatible with both the three-connected intermediate and cage compression mechanisms \cite{Alberti2017,Dusselier2017}.

Crystal graph isomorphism has three obvious limitations as a similarity metric for crystals: (i) it is a binary metric, thus unable to assign intermediate similarity values for different graphs; (ii) it is a computationally expensive test for large graphs; and (iii) it is not invariant to the choice of the unit cell. To address (i) and (ii), we adapted the D-measure \cite{Schieber2017} for multigraphs. This similarity compares graph connectivities based on distributions of node distances, generating a continuous metric space which recovers graph isomorphism with zero distance\cite{Schieber2017}. Then, a variational approach is proposed to solve (iii). When comparing a zeolite pair, we search for two supercells with equal number of T atoms and minimum discrepancy in terms of lattice geometry. This allows the comparison between crystals with different point groups, as is the case of most transformations in the literature (Supp. Tab. I), in a generalization of the coincidence lattice method for surfaces \cite{Koda2016} with a custom graph metric (see Methods). The topological distance between two crystals is then taken as the D-measure between the best-matched supercell graphs (Figs. \ref{fig:graph}b,c). Furthermore, to investigate the role of 3D atomic arrangement, we combined structural similarities under a single descriptor by using the Smooth Overlap of Atomic Positions (SOAP) approach \cite{Bartok2013}. We observed that kernel distance is well correlated with framework density and with CBUs (see Suppl. Figs. 3-6).

\begin{figure*}[tb]
\includegraphics[width=\linewidth]{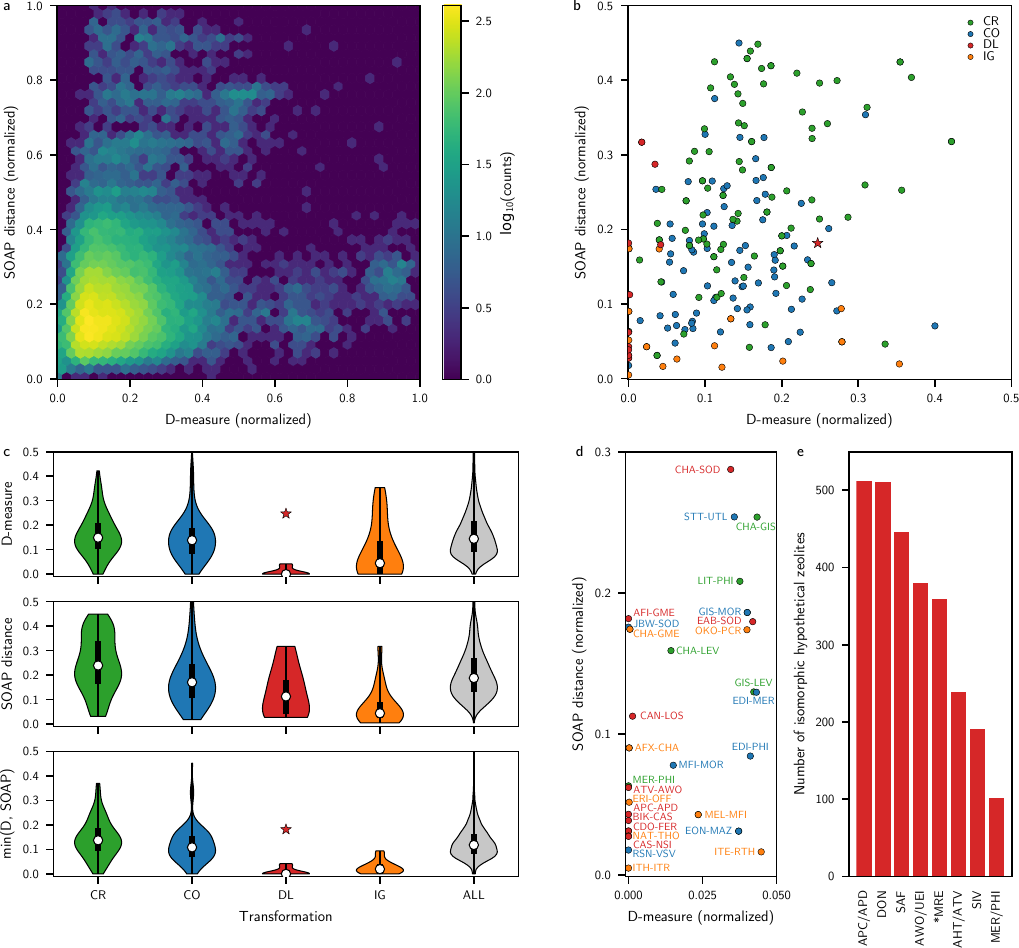}
\caption{\textbf{Structural and graph similarities in zeolites.} \textbf{a}, Distribution of zeolite pairs according to their (normalized) SOAP and graph distances. \textbf{b}, Distribution of experimentally known zeolite transformations in the joint SOAP-graph space and \textbf{c}, separately. Diffusionless transformations (DL) and intergrowth (IG) have smaller graph distances when compared to recrystallization (CR) and competing (CO) processes, as also evidenced by \textbf{d}. The only exception is the LTA-IFY transformation \cite{Jorda2013Synthesis}, indicated by a star. In \textbf{c}, the white dot indicate the median of the distribution. The bottom and top of the thicker solid lines are the first and third quartiles, respectively. The whiskers indicate the maximum and the minimum of the range. As opposed to similarity in density or in CBUs, high topological similarity is rare, as evidenced by the narrow base of the ALL violin plots. \textbf{e}, Histogram of the number of hypothetical zeolites which are graph isomorphic to known frameworks. Only the eight most frequent codes are shown in the figure, sorted in descending order of frequency.}
\label{fig:soap_d}
\end{figure*}

Fig. \ref{fig:soap_d}a illustrates the distribution of the pairwise distances between zeolites according to the two descriptors. When plotting all known interzeolite synthesis relations from our database, all diffusionless transformations fall in the lower graph distance region (Figs. \ref{fig:soap_d}b-d). This advances the argument that graph similarity is a powerful predictor of kinetic transformations in zeolites. The only exception to this rule is the reconstructive LTA-IFY transformation \cite{Jorda2013Synthesis}, which reportedly requires an external pressure of 3 GPa to be induced. This very high pressure suggests that diffusionless transitions between low graph-similar zeolites require extreme conditions. A large number of intergrown frameworks also occur for low graph-distance pairs. Interestingly, the SOAP descriptor complements the predictions of zeolite intergrowth, by capturing intergrown pairs with high structural similarity. This indicates the intergrowth may be possible if: the two polymorphs have similar atomic connectivities, which allow nucleating them to interconnect on similar surfaces upon nucleation (low graph distance); or if they have similar building units, through which the stacking process can be formed (low SOAP distance). For recrystallization transformations and competing phases, the majority of the pairs have higher dissimilarities across both metrics. The high predictive power of the proposed metrics can be combined with the kinetic understanding of experimental processes to distinguish diffusionless transformations. 

Since the topological similarity between two structures is a strong indicator that they may be related by intergrowth or diffusionless transformations, new frameworks could be accessed from known ones as parent structures. We looked for isomorphic pairs in hypothetical zeolite databases \cite{Deem2009ComputationalMaterials, Pophale2011AMaterials} which have energy above quartz. When all pairwise combinations of 269,515 enumerated structures and the known zeolites are assessed with the isomorphism criterion, 3,879 different hypothetical frameworks are found to have at least one known isomorphic counterpart. Based on previous evidence of accessibility, we suggest these structures would be strong candidates for the synthesis of novel zeolites. This list can be further refined according to their potential applications\cite{Lin2012InMaterials}, novelty, or availability of their parent frameworks. Fig. \ref{fig:soap_d}e shows how many hypothetical zeolites relate to known ones based on frequency of isomorphism. A high number of polymorphs are graph isomorphic with APC/APD, DON, SAF, AWO/UEI, *MRE, ATV/AHT and SIV. This may be related to the richness of graph-equivalent topologies that can be constructed using an orthorhombic unit cell within the search parameters used by Deem \textit{et al}.\cite{Deem2009ComputationalMaterials} to design the zeolite unit cells. Moreover, most of these crystals are better characterized by chains instead of just building blocks, whose regularity may play a role in the definition of the graph topology.

Analogously, we propose new diffusionless transformations for known zeolites given their graph similarity and new intergrown zeolites using the combination of topological and structural descriptors. 237 pairs of known zeolites have normalized graph distance under 0.01 (Supp. Tab. II), and 136 have SOAP distance smaller or equal to 0.04 (Supp. Tab. III). This tight threshold captures 60\% of all pairs related by diffusionless transformations and intergrowth. These pairs would be the best starting points for further experimental verification of graph-based relationships and to pursue new synthesis routes or intergrown frameworks with desired applications and chemical compositions. Furthermore, leveraging differences in space groups between graph-similar pairs could enable martensitic transformations using anisotropic stresses \cite{Zeng2016CrystalCeramics}. Finally, based on the distribution of distances between zeolites, we can determine which known frameworks are unique in terms of structure and topology. Supp. Figs. 9 and 10 show the distribution of pairwise distances between every known IZA framework and enumerated hypothetical zeolites. Whereas SOAP distances identify zeolites with lower framework densities (see Supp. Fig. 6, 9), the graph distance also detects exotic topologies through pairwise comparisons (Supplementary Fig. 10).

In summary, we have developed a computational approach based on graph theory to explain and predict interzeolite transformations. We performed a broad computationally-assisted mining of existing literature to clarify prevalent mechanisms of crystal phase relations. The extracted data was analyzed with structural and graph similarity metrics. We show how typical descriptors cannot explain pairwise interactions between zeolite crystals. Topological similarity provides an augmented complementary, quantitative metric to understand zeolite transformations. We employ the graph isomorphism test to explain diffusionless interzeolite conversions regardless of their building units. By generalizing this comparison to a topological similarity applied over supercells, we obtain remarkable agreement with the experimental results retrieved from the literature. The metrics predict hundreds of pairs of known zeolites with promising similarity and thousands of hypothetical frameworks which can have experimental relationships with known topologies. Predicted pairwise relationships can lead to the assembly of frameworks with different compositions and open new pathways for OSDA-free synthesis. Furthermore, the method for determining zeolite similarities is atom-agnostic, and could be extended to other systems, such as metal-organic frameworks, covalent-organic frameworks, as well as other systems related by diffusionless transformations.

\section*{Acknowledgements}
D.S.-K. acknowledges the MIT Nicole and Ingo Wender Fellowship and the MIT Robert Rose Presidential Fellowship for financial support. R.G.-B. thanks MIT DMSE and Toyota Faculty Chair for support. The work of E.O. and Z.J. was partially funded by the National Science Foundation Award \#1534340, DMREF, and the Office of Naval Research (ONR) under Contract No. N00014-16-1-2432. D.S.-K. and R.G.-B. thank A. Corma, M. Moliner and Y. Roman for fruitful discussions.

\section*{Author Contributions}
R.G.-B. conceived the project. D.S.-K. and R.G.-B. formulated the hypothesis of graph-similar transformations. D.S.-K. developed the graph and supercell matching methods, wrote their code, and performed all calculations. Z.J. and E.O. performed the literature mining and database query. Z.J., E.O. and D.S.-K. reviewed the extracted articles. D.S.-K. and R.G.-B. wrote the first version of the manuscript and made the figures. All authors contributed to the Methods section and to the final version of the manuscript.

\section*{Competing Interests}
The authors declare no competing interests.

\section*{References}

\clearpage
\section*{Methods}

\noindent \textbf{Literature extraction} Papers describing interzeolite transitions were found from a database of over 2 million material science and chemistry articles using a natural language processing (NLP) pipeline\cite{Kim2017Machine-learnedMaterials}. First, a subsection of approximately 70,000 papers related to zeolite materials was extracted by searching for variants of the word ``zeolite'' within the text of each paper. Then, papers containing interzeolite transitions were found by searching for pairs of zeolite materials with different IZA structures within the title, abstract, and synthesis paragraphs of the paper, as determined by the NLP pipeline\cite{kim2018inorganic}. These sections were also searched for keywords relating to interzeolite transitions such as ``inter-*'', ``intergrowth'', ``topological'', ``reconstruction'', and ``ADOR''. More than 540 papers contained at least one pair of structures and at least one keyword match. They were manually checked to determine the type of transition and pairs of zeolites involved in the inter-zeolite transition.
\\

\noindent \textbf{Database of zeolites} A database of 245 known zeolite frameworks was downloaded from the Database of Zeolite Structures, kept by the Structure Commission of the International Zeolite Association (IZA) \cite{IZADatabase}, including interrupted and partially disordered frameworks. Their framework density and CBUs was also extracted from the website. We selected the hypothetical zeolite database generated and optimized using the Sanders-Leslie-Catlow (SLC) force field \cite{Schroder1992} by Deem \textit{et al}. \cite{Deem2009ComputationalMaterials, Pophale2011AMaterials}. From the complete database with about 314,000 entries, we removed those whose energies were lower than quartz, ending up with 269,515 zeolite structures. All zeolites are considered in their pure silicate form.
\\

\noindent \textbf{Construction of zeolite multigraphs} We adopted a multigraph representation of crystals satisfying periodic boundary conditions \cite{Xie2018} for pure zeolite silicate structures. To maximize the amount of information embedded in the graph while minimizing its size, we removed the oxygen atoms from the graph. Each crystal graph contains as many nodes as Si atoms in its input unit cell. Nodes are connected if and only if their correspondent Si atoms share an oxygen atom. This avoids errors from the usage of nearest-neighbors search using Voronoi diagrams for porous materials. No node or edge labels are included in the graph.
\\

\noindent \textbf{Comparing crystal graphs} Multigraph isomorphism is performed using the VF2 algorithm \cite{Cordella2004} as implemented at NetworkX \cite{SciPyProceedings_11}. The graph similarity D-measure is implemented as reported by Schieber \textit{et al}. \cite{Schieber2017}. As the graph complement cannot be computed for multigraphs, the alpha-centrality is taken with respect to the graph without parallel edges. This still preserves notions of connectivities and distances.

The variational approach for the graph similarity starts with two crystals A, B with lattice matrices

\begin{align}\label{eq:lattices}
A &= \begin{bmatrix}
\mathbf{a}_1 & \mathbf{a}_2 & \mathbf{a}_3\\
\end{bmatrix}, \\
B &= \begin{bmatrix}
\mathbf{b}_1 & \mathbf{b}_2 & \mathbf{b}_3\\
\end{bmatrix},
\end{align}

\noindent with $\mathbf{a}_i, \mathbf{b}_j$ lattice vectors associated to each of the crystals. Each unit cell given by $A$, $B$ contains a number $n_T^{(A, B)}$ of T atoms in it. Given the lattice and the atomic positions, we can construct a graph $G_{X} = g(X)$ for the crystal $X$ using the crystal graph constructor $g: \mathcal{C} \rightarrow \mathcal{G}$. This maps the space of crystal structures (made from lattice and atomic basis) $\mathcal{C}$ to the space of graphs $\mathcal{G}$. As demonstrated by the isomorphism, the constructor $g$ is not necessarily injective.

Ultimately, we want to compare the topology between two graphs with the same number of nodes. This determines a scaling $m^{(A, B)}$ necessary to equalize the mass of both supercells,

\begin{align}\label{eq:scaling}
m^{(A,B)} &= \frac{\lcm \left(n_T^{(A)}, n_T^{(B)}\right)}{n_T^{(A,B)}},
\end{align}

\noindent where $\mathrm{LCM} \left(n_T^{(A)}, n_T^{(B)}\right)$ is the least common multiple between $n_T^{(A)}$ and $n_T^{(B)}$. We can then look for the transformation matrices $M^{(A, B)}$ that minimize the graph distance $d_\mathcal{G}(G_A, G_B)$ between two graphs $G_A, G_B$ created using the function $g$,

\begin{align}\label{eq:distance}
d(A, B) &= \min_{M^{(A)}, M^{(B)}} d_\mathcal{G}\left(g\left(M^{(A)} A\right), g\left(M^{(B)} B\right)\right),
\end{align}

\noindent subject to the constraints

\begin{align}\label{eq:constraint}
\det M^{(A, B)} &= m^{(A, B)}, \\
m_{ij}^{(A, B)} &\in \mathbb{Z}.
\end{align}

\noindent The resulting supercells are described by the matrices $M^{(A, B)}$. The crystal A, B supercells have lattice matrices given by $M^{(A)}A$, $M^{(B)}B$ plus the resulting atomic basis. A practical implementation is described in the Supplementary Information.
\\

\noindent \textbf{SOAP} For each atomic environment $\mathcal{X}_i$ in the crystal structure, a SOAP power spectrum $\mathbf{p}(\mathcal{X}_i)$ is calculated \cite{Bartok2013, Jager2018} using $r_\textrm{cut} = 10$, radial basis size of 8 with $L_\textrm{max} = 5$ as hyperparameters. Instead of adopting the regularized entropy match kernel \cite{De2016} to compare crystal structures, we opt for the average SOAP fingerprint given by

\begin{equation}\label{eq:soap_power}
    \mathbf{p}(Z) = \frac{1}{N} \sum_i \mathbf{p} (\mathcal{X}_i^Z)
\end{equation}

\noindent for each crystal structure $Z$. This allow us to simplify the analysis to hundreds of thousands of zeolites with varying number of atoms.

\noindent The unnormalized average structure kernel $K$ can be defined as the inner product of their power spectra,\cite{De2016}

\begin{equation}\label{eq:soap_kernel}
    K (A, B) = \mathbf{p}(A) \cdot \mathbf{p}(B),
\end{equation}

\noindent leading to the normalized kernel equivalent to a cosine similarity,

\begin{equation}\label{eq:soap_kernel_norm}
    \bar{K} (A, B) = \frac{K (A, B)}{\sqrt{K (A, A) K (B, B)}}.
\end{equation}

\noindent This kernel induces a metric $d_\textrm{kernel}$ given by \cite{De2016}

\begin{equation}\label{eq:soap_kernel_dist}
     d_\textrm{kernel}(A, B) = \sqrt{2 - 2 \bar{K} (A, B)}.
\end{equation}


\begin{thebibliography}{36}%
\makeatletter
\providecommand \@ifxundefined [1]{%
 \@ifx{#1\undefined}
}%
\providecommand \@ifnum [1]{%
 \ifnum #1\expandafter \@firstoftwo
 \else \expandafter \@secondoftwo
 \fi
}%
\providecommand \@ifx [1]{%
 \ifx #1\expandafter \@firstoftwo
 \else \expandafter \@secondoftwo
 \fi
}%
\providecommand \natexlab [1]{#1}%
\providecommand \enquote  [1]{``#1''}%
\providecommand \bibnamefont  [1]{#1}%
\providecommand \bibfnamefont [1]{#1}%
\providecommand \citenamefont [1]{#1}%
\providecommand \href@noop [0]{\@secondoftwo}%
\providecommand \href [0]{\begingroup \@sanitize@url \@href}%
\providecommand \@href[1]{\@@startlink{#1}\@@href}%
\providecommand \@@href[1]{\endgroup#1\@@endlink}%
\providecommand \@sanitize@url [0]{\catcode `\\12\catcode `\$12\catcode
  `\&12\catcode `\#12\catcode `\^12\catcode `\_12\catcode `\%12\relax}%
\providecommand \@@startlink[1]{}%
\providecommand \@@endlink[0]{}%
\providecommand \url  [0]{\begingroup\@sanitize@url \@url }%
\providecommand \@url [1]{\endgroup\@href {#1}{\urlprefix }}%
\providecommand \urlprefix  [0]{URL }%
\providecommand \Eprint [0]{\href }%
\providecommand \doibase [0]{http://dx.doi.org/}%
\providecommand \selectlanguage [0]{\@gobble}%
\providecommand \bibinfo  [0]{\@secondoftwo}%
\providecommand \bibfield  [0]{\@secondoftwo}%
\providecommand \translation [1]{[#1]}%
\providecommand \BibitemOpen [0]{}%
\providecommand \bibitemStop [0]{}%
\providecommand \bibitemNoStop [0]{.\EOS\space}%
\providecommand \EOS [0]{\spacefactor3000\relax}%
\providecommand \BibitemShut  [1]{\csname bibitem#1\endcsname}%
\let\auto@bib@innerbib\@empty
\bibitem [{\citenamefont {Davis}(2002)}]{Davis2002OrderedApplications}%
  \BibitemOpen
  \bibfield  {author} {\bibinfo {author} {\bibfnamefont {M.~E.}\ \bibnamefont
  {Davis}},\ }\bibfield  {title} {\enquote {\bibinfo {title} {{Ordered porous
  materials for emerging applications}},}\ }\href {\doibase
  10.1038/nature00785} {\bibfield  {journal} {\bibinfo  {journal} {Nature}\
  }\textbf {\bibinfo {volume} {417}},\ \bibinfo {pages} {813--821} (\bibinfo
  {year} {2002})}\BibitemShut {NoStop}%
\bibitem [{\citenamefont {Maldonado}\ \emph {et~al.}(2013)\citenamefont
  {Maldonado}, \citenamefont {Oleksiak}, \citenamefont {Chinta},\ and\
  \citenamefont {Rimer}}]{Maldonado2013}%
  \BibitemOpen
  \bibfield  {author} {\bibinfo {author} {\bibfnamefont {M.}~\bibnamefont
  {Maldonado}}, \bibinfo {author} {\bibfnamefont {M.~D.}\ \bibnamefont
  {Oleksiak}}, \bibinfo {author} {\bibfnamefont {S.}~\bibnamefont {Chinta}}, \
  and\ \bibinfo {author} {\bibfnamefont {J.~D.}\ \bibnamefont {Rimer}},\
  }\bibfield  {title} {\enquote {\bibinfo {title} {{Controlling crystal
  polymorphism in organic-free synthesis of Na-zeolites}},}\ }\href {\doibase
  10.1021/ja3105939} {\bibfield  {journal} {\bibinfo  {journal} {Journal of the
  American Chemical Society}\ }\textbf {\bibinfo {volume} {135}},\ \bibinfo
  {pages} {2641--2652} (\bibinfo {year} {2013})}\BibitemShut {NoStop}%
\bibitem [{\citenamefont {Li}, \citenamefont {Corma},\ and\ \citenamefont
  {Yu}(2015)}]{Li2015SynthesisStructures}%
  \BibitemOpen
  \bibfield  {author} {\bibinfo {author} {\bibfnamefont {J.}~\bibnamefont
  {Li}}, \bibinfo {author} {\bibfnamefont {A.}~\bibnamefont {Corma}}, \ and\
  \bibinfo {author} {\bibfnamefont {J.}~\bibnamefont {Yu}},\ }\bibfield
  {title} {\enquote {\bibinfo {title} {{Synthesis of new zeolite
  structures}},}\ }\href {\doibase 10.1039/c5cs00023h} {\bibfield  {journal}
  {\bibinfo  {journal} {Chemical Society Reviews}\ }\textbf {\bibinfo {volume}
  {44}},\ \bibinfo {pages} {7112--7127} (\bibinfo {year} {2015})}\BibitemShut
  {NoStop}%
\bibitem [{\citenamefont {Gallego}\ \emph {et~al.}(2017)\citenamefont
  {Gallego}, \citenamefont {Portilla}, \citenamefont {Paris}, \citenamefont
  {Le{\'{o}}n-Escamilla}, \citenamefont {Boronat}, \citenamefont {Moliner},\
  and\ \citenamefont {Corma}}]{Gallego2017AbReactions}%
  \BibitemOpen
  \bibfield  {author} {\bibinfo {author} {\bibfnamefont {E.~M.}\ \bibnamefont
  {Gallego}}, \bibinfo {author} {\bibfnamefont {M.~T.}\ \bibnamefont
  {Portilla}}, \bibinfo {author} {\bibfnamefont {C.}~\bibnamefont {Paris}},
  \bibinfo {author} {\bibfnamefont {A.}~\bibnamefont {Le{\'{o}}n-Escamilla}},
  \bibinfo {author} {\bibfnamefont {M.}~\bibnamefont {Boronat}}, \bibinfo
  {author} {\bibfnamefont {M.}~\bibnamefont {Moliner}}, \ and\ \bibinfo
  {author} {\bibfnamefont {A.}~\bibnamefont {Corma}},\ }\bibfield  {title}
  {\enquote {\bibinfo {title} {{"Ab initio" synthesis of zeolites for
  preestablished catalytic reactions}},}\ }\href {\doibase
  10.1126/science.aal0121} {\bibfield  {journal} {\bibinfo  {journal}
  {Science}\ }\textbf {\bibinfo {volume} {355}},\ \bibinfo {pages} {1051--1054}
  (\bibinfo {year} {2017})}\BibitemShut {NoStop}%
\bibitem [{\citenamefont {Honda}\ \emph {et~al.}(2013)\citenamefont {Honda},
  \citenamefont {Itakura}, \citenamefont {Matsuura}, \citenamefont {Onda},
  \citenamefont {Ide}, \citenamefont {Sadakane},\ and\ \citenamefont
  {Sano}}]{Honda2013RoleProcess.}%
  \BibitemOpen
  \bibfield  {author} {\bibinfo {author} {\bibfnamefont {K.}~\bibnamefont
  {Honda}}, \bibinfo {author} {\bibfnamefont {M.}~\bibnamefont {Itakura}},
  \bibinfo {author} {\bibfnamefont {Y.}~\bibnamefont {Matsuura}}, \bibinfo
  {author} {\bibfnamefont {A.}~\bibnamefont {Onda}}, \bibinfo {author}
  {\bibfnamefont {Y.}~\bibnamefont {Ide}}, \bibinfo {author} {\bibfnamefont
  {M.}~\bibnamefont {Sadakane}}, \ and\ \bibinfo {author} {\bibfnamefont
  {T.}~\bibnamefont {Sano}},\ }\bibfield  {title} {\enquote {\bibinfo {title}
  {{Role of structural similarity between starting zeolite and product zeolite
  in the interzeolite conversion process.}}}\ }\href
  {http://www.ncbi.nlm.nih.gov/pubmed/23763196} {\bibfield  {journal} {\bibinfo
   {journal} {Journal of Nanoscience and Nanotechnology}\ }\textbf {\bibinfo
  {volume} {13}},\ \bibinfo {pages} {3020--6} (\bibinfo {year}
  {2013})}\BibitemShut {NoStop}%
\bibitem [{\citenamefont {Marler}\ and\ \citenamefont
  {Gies}(2012)}]{Marler2012HydrousReview}%
  \BibitemOpen
  \bibfield  {author} {\bibinfo {author} {\bibfnamefont {B.}~\bibnamefont
  {Marler}}\ and\ \bibinfo {author} {\bibfnamefont {H.}~\bibnamefont {Gies}},\
  }\bibfield  {title} {\enquote {\bibinfo {title} {{Hydrous layer silicates as
  precursors for zeolites obtained through topotactic condensation: a
  review}},}\ }\href {\doibase 10.1127/0935-1221/2012/0024-2187} {\bibfield
  {journal} {\bibinfo  {journal} {European Journal of Mineralogy}\ }\textbf
  {\bibinfo {volume} {24}},\ \bibinfo {pages} {405--428} (\bibinfo {year}
  {2012})}\BibitemShut {NoStop}%
\bibitem [{\citenamefont {Eli{\'{a}}{\v{s}}ov{\'{a}}}\ \emph
  {et~al.}(2015)\citenamefont {Eli{\'{a}}{\v{s}}ov{\'{a}}}, \citenamefont
  {Opanasenko}, \citenamefont {Wheatley}, \citenamefont {Shamzhy},
  \citenamefont {Mazur}, \citenamefont {Nachtigall}, \citenamefont {Roth},
  \citenamefont {Morris},\ and\ \citenamefont
  {{\v{C}}ejka}}]{Eliasova2015TheZeolites}%
  \BibitemOpen
  \bibfield  {author} {\bibinfo {author} {\bibfnamefont {P.}~\bibnamefont
  {Eli{\'{a}}{\v{s}}ov{\'{a}}}}, \bibinfo {author} {\bibfnamefont
  {M.}~\bibnamefont {Opanasenko}}, \bibinfo {author} {\bibfnamefont {P.~S.}\
  \bibnamefont {Wheatley}}, \bibinfo {author} {\bibfnamefont {M.}~\bibnamefont
  {Shamzhy}}, \bibinfo {author} {\bibfnamefont {M.}~\bibnamefont {Mazur}},
  \bibinfo {author} {\bibfnamefont {P.}~\bibnamefont {Nachtigall}}, \bibinfo
  {author} {\bibfnamefont {W.~J.}\ \bibnamefont {Roth}}, \bibinfo {author}
  {\bibfnamefont {R.~E.}\ \bibnamefont {Morris}}, \ and\ \bibinfo {author}
  {\bibfnamefont {J.}~\bibnamefont {{\v{C}}ejka}},\ }\bibfield  {title}
  {\enquote {\bibinfo {title} {{The ADOR mechanism for the synthesis of new
  zeolites}},}\ }\href {\doibase 10.1039/C5CS00045A} {\bibfield  {journal}
  {\bibinfo  {journal} {Chemical Society Reviews}\ }\textbf {\bibinfo {volume}
  {44}},\ \bibinfo {pages} {7177--7206} (\bibinfo {year} {2015})}\BibitemShut
  {NoStop}%
\bibitem [{\citenamefont {Li}, \citenamefont {Moliner},\ and\ \citenamefont
  {Corma}(2018)}]{Li2018}%
  \BibitemOpen
  \bibfield  {author} {\bibinfo {author} {\bibfnamefont {C.}~\bibnamefont
  {Li}}, \bibinfo {author} {\bibfnamefont {M.}~\bibnamefont {Moliner}}, \ and\
  \bibinfo {author} {\bibfnamefont {A.}~\bibnamefont {Corma}},\ }\bibfield
  {title} {\enquote {\bibinfo {title} {{Building Zeolites from Precrystallized
  Units: Nanoscale Architecture}},}\ }\href {\doibase 10.1002/anie.201711422}
  {\bibfield  {journal} {\bibinfo  {journal} {Angewandte Chemie International
  Edition}\ } (\bibinfo {year} {2018}),\ 10.1002/anie.201711422}\BibitemShut
  {NoStop}%
\bibitem [{\citenamefont {Goel}, \citenamefont {Zones},\ and\ \citenamefont
  {Iglesia}(2015)}]{Goel2015SynthesisAgents}%
  \BibitemOpen
  \bibfield  {author} {\bibinfo {author} {\bibfnamefont {S.}~\bibnamefont
  {Goel}}, \bibinfo {author} {\bibfnamefont {S.~I.}\ \bibnamefont {Zones}}, \
  and\ \bibinfo {author} {\bibfnamefont {E.}~\bibnamefont {Iglesia}},\
  }\bibfield  {title} {\enquote {\bibinfo {title} {{Synthesis of Zeolites via
  Interzeolite Transformations without Organic Structure-Directing Agents}},}\
  }\href {\doibase 10.1021/cm504510f} {\bibfield  {journal} {\bibinfo
  {journal} {Chemistry of Materials}\ }\textbf {\bibinfo {volume} {27}},\
  \bibinfo {pages} {2056--2066} (\bibinfo {year} {2015})}\BibitemShut {NoStop}%
\bibitem [{\citenamefont {Baerlocher}, \citenamefont {McCusker},\ and\
  \citenamefont {Olson}(2007)}]{Baerlocher2007AtlasTypes}%
  \BibitemOpen
  \bibfield  {author} {\bibinfo {author} {\bibfnamefont {C.}~\bibnamefont
  {Baerlocher}}, \bibinfo {author} {\bibfnamefont {L.~B.}\ \bibnamefont
  {McCusker}}, \ and\ \bibinfo {author} {\bibfnamefont {D.~H.}\ \bibnamefont
  {Olson}},\ }\href@noop {} {\emph {\bibinfo {title} {{Atlas of Zeolite
  Framework Types}}}},\ \bibinfo {edition} {6th}\ ed.\ (\bibinfo  {publisher}
  {Elsevier Science},\ \bibinfo {address} {Amsterdam},\ \bibinfo {year}
  {2007})\ p.\ \bibinfo {pages} {404}\BibitemShut {NoStop}%
\bibitem [{\citenamefont {Zhang}, \citenamefont {Fernandez},\ and\
  \citenamefont {Ostraat}(2018)}]{Zhang2018Understanding}%
  \BibitemOpen
  \bibfield  {author} {\bibinfo {author} {\bibfnamefont {K.}~\bibnamefont
  {Zhang}}, \bibinfo {author} {\bibfnamefont {S.}~\bibnamefont {Fernandez}}, \
  and\ \bibinfo {author} {\bibfnamefont {M.~L.}\ \bibnamefont {Ostraat}},\
  }\bibfield  {title} {\enquote {\bibinfo {title} {{Understanding Commonalities
  and Interplay Between Organotemplate-Free Zeolite Synthesis, Hierarchical
  Structure Creation, and Interzeolite Transformation}},}\ }\href {\doibase
  10.1002/cctc.201800612} {\bibfield  {journal} {\bibinfo  {journal}
  {ChemCatChem}\ }\textbf {\bibinfo {volume} {10}},\ \bibinfo {pages}
  {4197--4212} (\bibinfo {year} {2018})}\BibitemShut {NoStop}%
\bibitem [{\citenamefont {O'Keeffe}\ and\ \citenamefont
  {Hyde}(1996)}]{OKeeffe1996TheStructures}%
  \BibitemOpen
  \bibfield  {author} {\bibinfo {author} {\bibfnamefont {M.}~\bibnamefont
  {O'Keeffe}}\ and\ \bibinfo {author} {\bibfnamefont {S.~T.}\ \bibnamefont
  {Hyde}},\ }\bibfield  {title} {\enquote {\bibinfo {title} {{The asymptotic
  behavior of coordination sequences for the 4-connected nets of zeolites and
  related structures}},}\ }\href {\doibase 10.1524/zkri.1996.211.2.73}
  {\bibfield  {journal} {\bibinfo  {journal} {Zeitschrift fur
  Kristallographie}\ }\textbf {\bibinfo {volume} {211}},\ \bibinfo {pages}
  {73--78} (\bibinfo {year} {1996})}\BibitemShut {NoStop}%
\bibitem [{\citenamefont {Foster}\ \emph {et~al.}(2004)\citenamefont {Foster},
  \citenamefont {Simperler}, \citenamefont {Bell}, \citenamefont {Friedrichs},
  \citenamefont {Paz},\ and\ \citenamefont {Klinowski}}]{Foster2004}%
  \BibitemOpen
  \bibfield  {author} {\bibinfo {author} {\bibfnamefont {M.~D.}\ \bibnamefont
  {Foster}}, \bibinfo {author} {\bibfnamefont {A.}~\bibnamefont {Simperler}},
  \bibinfo {author} {\bibfnamefont {R.~G.}\ \bibnamefont {Bell}}, \bibinfo
  {author} {\bibfnamefont {O.~D.}\ \bibnamefont {Friedrichs}}, \bibinfo
  {author} {\bibfnamefont {F.~A.~A.}\ \bibnamefont {Paz}}, \ and\ \bibinfo
  {author} {\bibfnamefont {J.}~\bibnamefont {Klinowski}},\ }\bibfield  {title}
  {\enquote {\bibinfo {title} {{Chemically feasible hypothetical crystalline
  networks}},}\ }\href {\doibase 10.1038/nmat1090} {\bibfield  {journal}
  {\bibinfo  {journal} {Nature Materials}\ }\textbf {\bibinfo {volume} {3}},\
  \bibinfo {pages} {234--238} (\bibinfo {year} {2004})}\BibitemShut {NoStop}%
\bibitem [{\citenamefont {Treacy}\ \emph {et~al.}(2004)\citenamefont {Treacy},
  \citenamefont {Rivin}, \citenamefont {Balkovsky}, \citenamefont {Randall},\
  and\ \citenamefont {Foster}}]{Treacy2004EnumerationGraphs}%
  \BibitemOpen
  \bibfield  {author} {\bibinfo {author} {\bibfnamefont {M.}~\bibnamefont
  {Treacy}}, \bibinfo {author} {\bibfnamefont {I.}~\bibnamefont {Rivin}},
  \bibinfo {author} {\bibfnamefont {E.}~\bibnamefont {Balkovsky}}, \bibinfo
  {author} {\bibfnamefont {K.}~\bibnamefont {Randall}}, \ and\ \bibinfo
  {author} {\bibfnamefont {M.}~\bibnamefont {Foster}},\ }\bibfield  {title}
  {\enquote {\bibinfo {title} {{Enumeration of periodic tetrahedral frameworks.
  II. Polynodal graphs}},}\ }\href {\doibase 10.1016/j.micromeso.2004.06.013}
  {\bibfield  {journal} {\bibinfo  {journal} {Microporous and Mesoporous
  Materials}\ }\textbf {\bibinfo {volume} {74}},\ \bibinfo {pages} {121--132}
  (\bibinfo {year} {2004})}\BibitemShut {NoStop}%
\bibitem [{\citenamefont {Witman}\ \emph {et~al.}(2018)\citenamefont {Witman},
  \citenamefont {Ling}, \citenamefont {Boyd}, \citenamefont {Barthel},
  \citenamefont {Haranczyk}, \citenamefont {Slater},\ and\ \citenamefont
  {Smit}}]{Witman2018CuttingZeolites}%
  \BibitemOpen
  \bibfield  {author} {\bibinfo {author} {\bibfnamefont {M.}~\bibnamefont
  {Witman}}, \bibinfo {author} {\bibfnamefont {S.}~\bibnamefont {Ling}},
  \bibinfo {author} {\bibfnamefont {P.}~\bibnamefont {Boyd}}, \bibinfo {author}
  {\bibfnamefont {S.}~\bibnamefont {Barthel}}, \bibinfo {author} {\bibfnamefont
  {M.}~\bibnamefont {Haranczyk}}, \bibinfo {author} {\bibfnamefont
  {B.}~\bibnamefont {Slater}}, \ and\ \bibinfo {author} {\bibfnamefont
  {B.}~\bibnamefont {Smit}},\ }\bibfield  {title} {\enquote {\bibinfo {title}
  {{Cutting Materials in Half: A Graph Theory Approach for Generating Crystal
  Surfaces and Its Prediction of 2D Zeolites}},}\ }\href {\doibase
  10.1021/acscentsci.7b00555} {\bibfield  {journal} {\bibinfo  {journal} {ACS
  central science}\ }\textbf {\bibinfo {volume} {4}},\ \bibinfo {pages}
  {235--245} (\bibinfo {year} {2018})}\BibitemShut {NoStop}%
\bibitem [{\citenamefont {Cordella}\ \emph {et~al.}(2004)\citenamefont
  {Cordella}, \citenamefont {Foggia}, \citenamefont {Sansone},\ and\
  \citenamefont {Vento}}]{Cordella2004}%
  \BibitemOpen
  \bibfield  {author} {\bibinfo {author} {\bibfnamefont {L.~P.}\ \bibnamefont
  {Cordella}}, \bibinfo {author} {\bibfnamefont {P.}~\bibnamefont {Foggia}},
  \bibinfo {author} {\bibfnamefont {C.}~\bibnamefont {Sansone}}, \ and\
  \bibinfo {author} {\bibfnamefont {M.}~\bibnamefont {Vento}},\ }\bibfield
  {title} {\enquote {\bibinfo {title} {{A (sub)graph isomorphism algorithm for
  matching large graphs}},}\ }\href {\doibase 10.1109/TPAMI.2004.75} {\bibfield
   {journal} {\bibinfo  {journal} {IEEE Transactions on Pattern Analysis and
  Machine Intelligence}\ }\textbf {\bibinfo {volume} {26}},\ \bibinfo {pages}
  {1367--1372} (\bibinfo {year} {2004})}\BibitemShut {NoStop}%
\bibitem [{\citenamefont {Blatov}(2007)}]{Blatov2007}%
  \BibitemOpen
  \bibfield  {author} {\bibinfo {author} {\bibfnamefont {V.~A.}\ \bibnamefont
  {Blatov}},\ }\bibfield  {title} {\enquote {\bibinfo {title} {{Topological
  relations between three-dimensional periodic nets. I. Uninodal nets}},}\
  }\href {\doibase 10.1107/S0108767307022088} {\bibfield  {journal} {\bibinfo
  {journal} {Acta Crystallographica Section A: Foundations of Crystallography}\
  }\textbf {\bibinfo {volume} {63}},\ \bibinfo {pages} {329--343} (\bibinfo
  {year} {2007})}\BibitemShut {NoStop}%
\bibitem [{\citenamefont {Porter}, \citenamefont {Easterling},\ and\
  \citenamefont {Sherif}(2009)}]{porter2009phase}%
  \BibitemOpen
  \bibfield  {author} {\bibinfo {author} {\bibfnamefont {D.~A.}\ \bibnamefont
  {Porter}}, \bibinfo {author} {\bibfnamefont {K.~E.}\ \bibnamefont
  {Easterling}}, \ and\ \bibinfo {author} {\bibfnamefont {M.}~\bibnamefont
  {Sherif}},\ }\href@noop {} {\emph {\bibinfo {title} {{Phase Transformations
  in Metals and Alloys}}}},\ \bibinfo {edition} {3rd}\ ed.\ (\bibinfo
  {publisher} {CRC Press},\ \bibinfo {address} {Boca Raton},\ \bibinfo {year}
  {2009})\BibitemShut {NoStop}%
\bibitem [{\citenamefont {Alberti}, \citenamefont {Cruciani},\ and\
  \citenamefont {Martucci}(2017)}]{Alberti2017}%
  \BibitemOpen
  \bibfield  {author} {\bibinfo {author} {\bibfnamefont {A.}~\bibnamefont
  {Alberti}}, \bibinfo {author} {\bibfnamefont {G.}~\bibnamefont {Cruciani}}, \
  and\ \bibinfo {author} {\bibfnamefont {A.}~\bibnamefont {Martucci}},\
  }\bibfield  {title} {\enquote {\bibinfo {title} {{Reconstructive phase
  transitions induced by temperature in gmelinite-Na zeolite}},}\ }\href
  {\doibase 10.2138/am-2017-5910} {\bibfield  {journal} {\bibinfo  {journal}
  {American Mineralogist}\ }\textbf {\bibinfo {volume} {102}},\ \bibinfo
  {pages} {1727--1735} (\bibinfo {year} {2017})}\BibitemShut {NoStop}%
\bibitem [{\citenamefont {Dusselier}\ \emph {et~al.}(2017)\citenamefont
  {Dusselier}, \citenamefont {Kang}, \citenamefont {Xie},\ and\ \citenamefont
  {Davis}}]{Dusselier2017}%
  \BibitemOpen
  \bibfield  {author} {\bibinfo {author} {\bibfnamefont {M.}~\bibnamefont
  {Dusselier}}, \bibinfo {author} {\bibfnamefont {J.~H.}\ \bibnamefont {Kang}},
  \bibinfo {author} {\bibfnamefont {D.}~\bibnamefont {Xie}}, \ and\ \bibinfo
  {author} {\bibfnamefont {M.~E.}\ \bibnamefont {Davis}},\ }\bibfield  {title}
  {\enquote {\bibinfo {title} {{CIT-9: A Fault-Free Gmelinite Zeolite}},}\
  }\href {\doibase 10.1002/anie.201707452} {\bibfield  {journal} {\bibinfo
  {journal} {Angewandte Chemie - International Edition}\ }\textbf {\bibinfo
  {volume} {56}},\ \bibinfo {pages} {13475--13478} (\bibinfo {year}
  {2017})}\BibitemShut {NoStop}%
\bibitem [{\citenamefont {Schieber}\ \emph {et~al.}(2017)\citenamefont
  {Schieber}, \citenamefont {Carpi}, \citenamefont {D{\'{i}}az-Guilera},
  \citenamefont {Pardalos}, \citenamefont {Masoller},\ and\ \citenamefont
  {Ravetti}}]{Schieber2017}%
  \BibitemOpen
  \bibfield  {author} {\bibinfo {author} {\bibfnamefont {T.~A.}\ \bibnamefont
  {Schieber}}, \bibinfo {author} {\bibfnamefont {L.}~\bibnamefont {Carpi}},
  \bibinfo {author} {\bibfnamefont {A.}~\bibnamefont {D{\'{i}}az-Guilera}},
  \bibinfo {author} {\bibfnamefont {P.~M.}\ \bibnamefont {Pardalos}}, \bibinfo
  {author} {\bibfnamefont {C.}~\bibnamefont {Masoller}}, \ and\ \bibinfo
  {author} {\bibfnamefont {M.~G.}\ \bibnamefont {Ravetti}},\ }\bibfield
  {title} {\enquote {\bibinfo {title} {{Quantification of network structural
  dissimilarities}},}\ }\href {\doibase 10.1038/ncomms13928} {\bibfield
  {journal} {\bibinfo  {journal} {Nature Communications}\ }\textbf {\bibinfo
  {volume} {8}},\ \bibinfo {pages} {13928} (\bibinfo {year}
  {2017})}\BibitemShut {NoStop}%
\bibitem [{\citenamefont {Koda}\ \emph {et~al.}(2016)\citenamefont {Koda},
  \citenamefont {Bechstedt}, \citenamefont {Marques},\ and\ \citenamefont
  {Teles}}]{Koda2016}%
  \BibitemOpen
  \bibfield  {author} {\bibinfo {author} {\bibfnamefont {D.~S.}\ \bibnamefont
  {Koda}}, \bibinfo {author} {\bibfnamefont {F.}~\bibnamefont {Bechstedt}},
  \bibinfo {author} {\bibfnamefont {M.}~\bibnamefont {Marques}}, \ and\
  \bibinfo {author} {\bibfnamefont {L.~K.}\ \bibnamefont {Teles}},\ }\bibfield
  {title} {\enquote {\bibinfo {title} {{Coincidence Lattices of 2D Crystals:
  Heterostructure Predictions and Applications}},}\ }\href {\doibase
  10.1021/acs.jpcc.6b01496} {\bibfield  {journal} {\bibinfo  {journal} {Journal
  of Physical Chemistry C}\ }\textbf {\bibinfo {volume} {120}},\ \bibinfo
  {pages} {10895--10908} (\bibinfo {year} {2016})}\BibitemShut {NoStop}%
\bibitem [{\citenamefont {Bart{\'{o}}k}, \citenamefont {Kondor},\ and\
  \citenamefont {Cs{\'{a}}nyi}(2013)}]{Bartok2013}%
  \BibitemOpen
  \bibfield  {author} {\bibinfo {author} {\bibfnamefont {A.~P.}\ \bibnamefont
  {Bart{\'{o}}k}}, \bibinfo {author} {\bibfnamefont {R.}~\bibnamefont
  {Kondor}}, \ and\ \bibinfo {author} {\bibfnamefont {G.}~\bibnamefont
  {Cs{\'{a}}nyi}},\ }\bibfield  {title} {\enquote {\bibinfo {title} {{On
  representing chemical environments}},}\ }\href {\doibase
  10.1103/PhysRevB.87.184115} {\bibfield  {journal} {\bibinfo  {journal}
  {Physical Review B}\ }\textbf {\bibinfo {volume} {87}},\ \bibinfo {pages}
  {184115} (\bibinfo {year} {2013})}\BibitemShut {NoStop}%
\bibitem [{\citenamefont {Jord{\'{a}}}\ \emph {et~al.}(2013)\citenamefont
  {Jord{\'{a}}}, \citenamefont {Rey}, \citenamefont {Sastre}, \citenamefont
  {Valencia}, \citenamefont {Palomino}, \citenamefont {Corma}, \citenamefont
  {Segura}, \citenamefont {Errandonea}, \citenamefont {Lacomba}, \citenamefont
  {Manj{\'{o}}n}, \citenamefont {Gomis}, \citenamefont {Kleppe}, \citenamefont
  {Jephcoat}, \citenamefont {Amboage},\ and\ \citenamefont
  {Rodr{\'{i}}guez-Velamaz{\'{a}}n}}]{Jorda2013Synthesis}%
  \BibitemOpen
  \bibfield  {author} {\bibinfo {author} {\bibfnamefont {J.~L.}\ \bibnamefont
  {Jord{\'{a}}}}, \bibinfo {author} {\bibfnamefont {F.}~\bibnamefont {Rey}},
  \bibinfo {author} {\bibfnamefont {G.}~\bibnamefont {Sastre}}, \bibinfo
  {author} {\bibfnamefont {S.}~\bibnamefont {Valencia}}, \bibinfo {author}
  {\bibfnamefont {M.}~\bibnamefont {Palomino}}, \bibinfo {author}
  {\bibfnamefont {A.}~\bibnamefont {Corma}}, \bibinfo {author} {\bibfnamefont
  {A.}~\bibnamefont {Segura}}, \bibinfo {author} {\bibfnamefont
  {D.}~\bibnamefont {Errandonea}}, \bibinfo {author} {\bibfnamefont
  {R.}~\bibnamefont {Lacomba}}, \bibinfo {author} {\bibfnamefont {F.~J.}\
  \bibnamefont {Manj{\'{o}}n}}, \bibinfo {author} {\bibfnamefont
  {Ã.}~\bibnamefont {Gomis}}, \bibinfo {author} {\bibfnamefont {A.~K.}\
  \bibnamefont {Kleppe}}, \bibinfo {author} {\bibfnamefont {A.~P.}\
  \bibnamefont {Jephcoat}}, \bibinfo {author} {\bibfnamefont {M.}~\bibnamefont
  {Amboage}}, \ and\ \bibinfo {author} {\bibfnamefont {J.~A.}\ \bibnamefont
  {Rodr{\'{i}}guez-Velamaz{\'{a}}n}},\ }\bibfield  {title} {\enquote {\bibinfo
  {title} {{Synthesis of a Novel Zeolite through a Pressure-Induced
  Reconstructive Phase Transition Process}},}\ }\href {\doibase
  10.1002/anie.201305230} {\bibfield  {journal} {\bibinfo  {journal}
  {Angewandte Chemie International Edition}\ }\textbf {\bibinfo {volume}
  {52}},\ \bibinfo {pages} {10458--10462} (\bibinfo {year} {2013})}\BibitemShut
  {NoStop}%
\bibitem [{\citenamefont {Deem}\ \emph {et~al.}(2009)\citenamefont {Deem},
  \citenamefont {Pophale}, \citenamefont {Cheeseman},\ and\ \citenamefont
  {Earl}}]{Deem2009ComputationalMaterials}%
  \BibitemOpen
  \bibfield  {author} {\bibinfo {author} {\bibfnamefont {M.~W.}\ \bibnamefont
  {Deem}}, \bibinfo {author} {\bibfnamefont {R.}~\bibnamefont {Pophale}},
  \bibinfo {author} {\bibfnamefont {P.~A.}\ \bibnamefont {Cheeseman}}, \ and\
  \bibinfo {author} {\bibfnamefont {D.~J.}\ \bibnamefont {Earl}},\ }\bibfield
  {title} {\enquote {\bibinfo {title} {{Computational Discovery of New
  Zeolite-Like Materials}},}\ }\href {\doibase 10.1021/jp906984z} {\bibfield
  {journal} {\bibinfo  {journal} {The Journal of Physical Chemistry C}\
  }\textbf {\bibinfo {volume} {113}},\ \bibinfo {pages} {21353--21360}
  (\bibinfo {year} {2009})}\BibitemShut {NoStop}%
\bibitem [{\citenamefont {Pophale}, \citenamefont {Cheeseman},\ and\
  \citenamefont {Deem}(2011)}]{Pophale2011AMaterials}%
  \BibitemOpen
  \bibfield  {author} {\bibinfo {author} {\bibfnamefont {R.}~\bibnamefont
  {Pophale}}, \bibinfo {author} {\bibfnamefont {P.~A.}\ \bibnamefont
  {Cheeseman}}, \ and\ \bibinfo {author} {\bibfnamefont {M.~W.}\ \bibnamefont
  {Deem}},\ }\bibfield  {title} {\enquote {\bibinfo {title} {{A database of new
  zeolite-like materials}},}\ }\href {\doibase 10.1039/c0cp02255a} {\bibfield
  {journal} {\bibinfo  {journal} {Physical Chemistry Chemical Physics}\
  }\textbf {\bibinfo {volume} {13}},\ \bibinfo {pages} {12407--12412} (\bibinfo
  {year} {2011})}\BibitemShut {NoStop}%
\bibitem [{\citenamefont {Lin}\ \emph {et~al.}(2012)\citenamefont {Lin},
  \citenamefont {Berger}, \citenamefont {Martin}, \citenamefont {Kim},
  \citenamefont {Swisher}, \citenamefont {Jariwala}, \citenamefont {Rycroft},
  \citenamefont {Bhown}, \citenamefont {Deem}, \citenamefont {Haranczyk},\ and\
  \citenamefont {Smit}}]{Lin2012InMaterials}%
  \BibitemOpen
  \bibfield  {author} {\bibinfo {author} {\bibfnamefont {L.-C.}\ \bibnamefont
  {Lin}}, \bibinfo {author} {\bibfnamefont {A.~H.}\ \bibnamefont {Berger}},
  \bibinfo {author} {\bibfnamefont {R.~L.}\ \bibnamefont {Martin}}, \bibinfo
  {author} {\bibfnamefont {J.}~\bibnamefont {Kim}}, \bibinfo {author}
  {\bibfnamefont {J.~A.}\ \bibnamefont {Swisher}}, \bibinfo {author}
  {\bibfnamefont {K.}~\bibnamefont {Jariwala}}, \bibinfo {author}
  {\bibfnamefont {C.~H.}\ \bibnamefont {Rycroft}}, \bibinfo {author}
  {\bibfnamefont {A.~S.}\ \bibnamefont {Bhown}}, \bibinfo {author}
  {\bibfnamefont {M.~W.}\ \bibnamefont {Deem}}, \bibinfo {author}
  {\bibfnamefont {M.}~\bibnamefont {Haranczyk}}, \ and\ \bibinfo {author}
  {\bibfnamefont {B.}~\bibnamefont {Smit}},\ }\bibfield  {title} {\enquote
  {\bibinfo {title} {{In silico screening of carbon-capture materials}},}\
  }\href {\doibase 10.1038/nmat3336} {\bibfield  {journal} {\bibinfo  {journal}
  {Nature Materials}\ }\textbf {\bibinfo {volume} {11}},\ \bibinfo {pages}
  {633--641} (\bibinfo {year} {2012})}\BibitemShut {NoStop}%
\bibitem [{\citenamefont {Zeng}\ \emph {et~al.}(2016)\citenamefont {Zeng},
  \citenamefont {Lai}, \citenamefont {Gan},\ and\ \citenamefont
  {Schuh}}]{Zeng2016CrystalCeramics}%
  \BibitemOpen
  \bibfield  {author} {\bibinfo {author} {\bibfnamefont {X.~M.}\ \bibnamefont
  {Zeng}}, \bibinfo {author} {\bibfnamefont {A.}~\bibnamefont {Lai}}, \bibinfo
  {author} {\bibfnamefont {C.~L.}\ \bibnamefont {Gan}}, \ and\ \bibinfo
  {author} {\bibfnamefont {C.~A.}\ \bibnamefont {Schuh}},\ }\bibfield  {title}
  {\enquote {\bibinfo {title} {{Crystal orientation dependence of the
  stress-induced martensitic transformation in zirconia-based shape memory
  ceramics}},}\ }\href {\doibase 10.1016/J.ACTAMAT.2016.06.030} {\bibfield
  {journal} {\bibinfo  {journal} {Acta Materialia}\ }\textbf {\bibinfo {volume}
  {116}},\ \bibinfo {pages} {124--135} (\bibinfo {year} {2016})}\BibitemShut
  {NoStop}%
\bibitem [{\citenamefont {Kim}\ \emph {et~al.}(2017)\citenamefont {Kim},
  \citenamefont {Huang}, \citenamefont {Tomala}, \citenamefont {Matthews},
  \citenamefont {Strubell}, \citenamefont {Saunders}, \citenamefont
  {McCallum},\ and\ \citenamefont
  {Olivetti}}]{Kim2017Machine-learnedMaterials}%
  \BibitemOpen
  \bibfield  {author} {\bibinfo {author} {\bibfnamefont {E.}~\bibnamefont
  {Kim}}, \bibinfo {author} {\bibfnamefont {K.}~\bibnamefont {Huang}}, \bibinfo
  {author} {\bibfnamefont {A.}~\bibnamefont {Tomala}}, \bibinfo {author}
  {\bibfnamefont {S.}~\bibnamefont {Matthews}}, \bibinfo {author}
  {\bibfnamefont {E.}~\bibnamefont {Strubell}}, \bibinfo {author}
  {\bibfnamefont {A.}~\bibnamefont {Saunders}}, \bibinfo {author}
  {\bibfnamefont {A.}~\bibnamefont {McCallum}}, \ and\ \bibinfo {author}
  {\bibfnamefont {E.}~\bibnamefont {Olivetti}},\ }\bibfield  {title} {\enquote
  {\bibinfo {title} {{Machine-learned and codified synthesis parameters of
  oxide materials}},}\ }\href {\doibase 10.1038/sdata.2017.127} {\bibfield
  {journal} {\bibinfo  {journal} {Scientific data}\ }\textbf {\bibinfo {volume}
  {4}},\ \bibinfo {pages} {170127} (\bibinfo {year} {2017})}\BibitemShut
  {NoStop}%
\bibitem [{\citenamefont {Kim}\ \emph {et~al.}(2018)\citenamefont {Kim},
  \citenamefont {Jensen}, \citenamefont {van Grootel}, \citenamefont {Huang},
  \citenamefont {Staib}, \citenamefont {Mysore}, \citenamefont {Chang},
  \citenamefont {Strubell}, \citenamefont {McCallum}, \citenamefont {Jegelka},\
  and\ \citenamefont {Olivetti}}]{kim2018inorganic}%
  \BibitemOpen
  \bibfield  {author} {\bibinfo {author} {\bibfnamefont {E.}~\bibnamefont
  {Kim}}, \bibinfo {author} {\bibfnamefont {Z.}~\bibnamefont {Jensen}},
  \bibinfo {author} {\bibfnamefont {A.}~\bibnamefont {van Grootel}}, \bibinfo
  {author} {\bibfnamefont {K.}~\bibnamefont {Huang}}, \bibinfo {author}
  {\bibfnamefont {M.}~\bibnamefont {Staib}}, \bibinfo {author} {\bibfnamefont
  {S.}~\bibnamefont {Mysore}}, \bibinfo {author} {\bibfnamefont {H.-S.}\
  \bibnamefont {Chang}}, \bibinfo {author} {\bibfnamefont {E.}~\bibnamefont
  {Strubell}}, \bibinfo {author} {\bibfnamefont {A.}~\bibnamefont {McCallum}},
  \bibinfo {author} {\bibfnamefont {S.}~\bibnamefont {Jegelka}}, \ and\
  \bibinfo {author} {\bibfnamefont {E.}~\bibnamefont {Olivetti}},\ }\bibfield
  {title} {\enquote {\bibinfo {title} {{Inorganic Materials Synthesis Planning
  with Literature-Trained Neural Networks}},}\ }\href@noop {} {\bibfield
  {journal} {\bibinfo  {journal} {arXiv:1901.00032}\ } (\bibinfo {year}
  {2018})}\BibitemShut {NoStop}%
\bibitem [{\citenamefont {{Ch. Baerlocher and L.B.
  McCusker}}(2019)}]{IZADatabase}%
  \BibitemOpen
  \bibfield  {author} {\bibinfo {author} {\bibnamefont {{Ch. Baerlocher and
  L.B. McCusker}}},\ }\bibfield  {title} {\enquote {\bibinfo {title} {{Database
  of Zeolite Structures}},}\ }\href {http://www.iza-structure.org/databases/}
  {\bibfield  {journal} {\bibinfo  {journal}
  {http://www.iza-structure.org/databases/}\ } (\bibinfo {year}
  {2019})}\BibitemShut {NoStop}%
\bibitem [{\citenamefont {Schr{\"{o}}der}\ \emph {et~al.}(1992)\citenamefont
  {Schr{\"{o}}der}, \citenamefont {Sauer}, \citenamefont {Leslie},
  \citenamefont {Richard}, \citenamefont {Catlow},\ and\ \citenamefont
  {Thomas}}]{Schroder1992}%
  \BibitemOpen
  \bibfield  {author} {\bibinfo {author} {\bibfnamefont {K.~P.}\ \bibnamefont
  {Schr{\"{o}}der}}, \bibinfo {author} {\bibfnamefont {J.}~\bibnamefont
  {Sauer}}, \bibinfo {author} {\bibfnamefont {M.}~\bibnamefont {Leslie}},
  \bibinfo {author} {\bibfnamefont {C.}~\bibnamefont {Richard}}, \bibinfo
  {author} {\bibfnamefont {A.}~\bibnamefont {Catlow}}, \ and\ \bibinfo {author}
  {\bibfnamefont {J.~M.}\ \bibnamefont {Thomas}},\ }\bibfield  {title}
  {\enquote {\bibinfo {title} {{Bridging hydrodyl groups in zeolitic catalysts:
  a computer simulation of their structure, vibrational properties and acidity
  in protonated faujasites (HY zeolites)}},}\ }\href {\doibase
  10.1016/0009-2614(92)90030-Q} {\bibfield  {journal} {\bibinfo  {journal}
  {Chemical Physics Letters}\ }\textbf {\bibinfo {volume} {188}},\ \bibinfo
  {pages} {320--325} (\bibinfo {year} {1992})}\BibitemShut {NoStop}%
\bibitem [{\citenamefont {Xie}\ and\ \citenamefont {Grossman}(2018)}]{Xie2018}%
  \BibitemOpen
  \bibfield  {author} {\bibinfo {author} {\bibfnamefont {T.}~\bibnamefont
  {Xie}}\ and\ \bibinfo {author} {\bibfnamefont {J.~C.}\ \bibnamefont
  {Grossman}},\ }\bibfield  {title} {\enquote {\bibinfo {title} {{Crystal Graph
  Convolutional Neural Networks for an Accurate and Interpretable Prediction of
  Material Properties}},}\ }\href {\doibase 10.1103/PhysRevLett.120.145301}
  {\bibfield  {journal} {\bibinfo  {journal} {Physical Review Letters}\
  }\textbf {\bibinfo {volume} {120}},\ \bibinfo {pages} {145301} (\bibinfo
  {year} {2018})}\BibitemShut {NoStop}%
\bibitem [{\citenamefont {Hagberg}, \citenamefont {Schult},\ and\ \citenamefont
  {Swart}(2008)}]{SciPyProceedings_11}%
  \BibitemOpen
  \bibfield  {author} {\bibinfo {author} {\bibfnamefont {A.~A.}\ \bibnamefont
  {Hagberg}}, \bibinfo {author} {\bibfnamefont {D.~A.}\ \bibnamefont {Schult}},
  \ and\ \bibinfo {author} {\bibfnamefont {P.~J.}\ \bibnamefont {Swart}},\
  }\bibfield  {title} {\enquote {\bibinfo {title} {{Exploring Network
  Structure, Dynamics, and Function using NetworkX}},}\ }in\ \href@noop {}
  {\emph {\bibinfo {booktitle} {Proceedings of the 7th Python in Science
  Conference}}},\ \bibinfo {editor} {edited by\ \bibinfo {editor}
  {\bibfnamefont {G.}~\bibnamefont {Varoquaux}}, \bibinfo {editor}
  {\bibfnamefont {T.}~\bibnamefont {Vaught}}, \ and\ \bibinfo {editor}
  {\bibfnamefont {J.}~\bibnamefont {Millman}}}\ (\bibinfo {address} {Pasadena,
  CA USA},\ \bibinfo {year} {2008})\ pp.\ \bibinfo {pages} {11--15}\BibitemShut
  {NoStop}%
\bibitem [{\citenamefont {J{\"{a}}ger}\ \emph {et~al.}(2018)\citenamefont
  {J{\"{a}}ger}, \citenamefont {Morooka}, \citenamefont {Federici~Canova},
  \citenamefont {Himanen},\ and\ \citenamefont {Foster}}]{Jager2018}%
  \BibitemOpen
  \bibfield  {author} {\bibinfo {author} {\bibfnamefont {M.~O.~J.}\
  \bibnamefont {J{\"{a}}ger}}, \bibinfo {author} {\bibfnamefont {E.~V.}\
  \bibnamefont {Morooka}}, \bibinfo {author} {\bibfnamefont {F.}~\bibnamefont
  {Federici~Canova}}, \bibinfo {author} {\bibfnamefont {L.}~\bibnamefont
  {Himanen}}, \ and\ \bibinfo {author} {\bibfnamefont {A.~S.}\ \bibnamefont
  {Foster}},\ }\bibfield  {title} {\enquote {\bibinfo {title} {{Machine
  learning hydrogen adsorption on nanoclusters through structural
  descriptors}},}\ }\href {\doibase 10.1038/s41524-018-0096-5} {\bibfield
  {journal} {\bibinfo  {journal} {npj Computational Materials}\ }\textbf
  {\bibinfo {volume} {4}},\ \bibinfo {pages} {37} (\bibinfo {year}
  {2018})}\BibitemShut {NoStop}%
\bibitem [{\citenamefont {De}\ \emph {et~al.}(2016)\citenamefont {De},
  \citenamefont {Bart{\'{o}}k}, \citenamefont {Cs{\'{a}}nyi},\ and\
  \citenamefont {Ceriotti}}]{De2016}%
  \BibitemOpen
  \bibfield  {author} {\bibinfo {author} {\bibfnamefont {S.}~\bibnamefont
  {De}}, \bibinfo {author} {\bibfnamefont {A.~P.}\ \bibnamefont
  {Bart{\'{o}}k}}, \bibinfo {author} {\bibfnamefont {G.}~\bibnamefont
  {Cs{\'{a}}nyi}}, \ and\ \bibinfo {author} {\bibfnamefont {M.}~\bibnamefont
  {Ceriotti}},\ }\bibfield  {title} {\enquote {\bibinfo {title} {{Comparing
  molecules and solids across structural and alchemical space}},}\ }\href
  {\doibase 10.1039/C6CP00415F} {\bibfield  {journal} {\bibinfo  {journal}
  {Physical Chemistry Chemical Physics}\ }\textbf {\bibinfo {volume} {18}},\
  \bibinfo {pages} {13754--13769} (\bibinfo {year} {2016})}\BibitemShut
  {NoStop}%
\end{thebibliography}
\end{document}